\documentclass[showpacs,preprintnumbers]{revtex4}
\usepackage{amssymb}
\usepackage{amsmath}
\usepackage{graphicx}
\usepackage{dcolumn}
\usepackage{bm}
\usepackage{epsf}

\begin{document}

\title{Application of complex-scaling method for few-body scattering}
\author{Rimantas Lazauskas}
\email{rimantas.lazauskas@iphc.in2p3.fr}
\affiliation{IPHC, IN2P3-CNRS/Universit\'e Louis Pasteur BP 28, F-67037 Strasbourg Cedex
2, France}
\author{Jaume Carbonell}
\email{jaume.carbonellm@lpsc.in2p3.fr}
\affiliation{LPSC, 53 Av. des Martyrs, 38026 Grenoble, France}
\date{\today}

\begin{abstract}
Formalism based on complex-scaling method is developed for solving  the few particle
scattering problem by employing only trivial boundary conditions. Several applications
are presented proving efficiency of the method in describing elastic and
three-body break-up reactions for Hamiltonians which may include both short
and long-range interaction.
\end{abstract}

\maketitle












\section{Introduction}

The quantum-mechanical problem of interacting particles is of the fundamental
importance in theoretical physics, opening broad field of application related
with the description of the microworld.
Quick development of the computational
techniques, following the rapid evolution of the computational power,
provoked sizeable advance in multi-particle bound state problem: rigorous
and accurate description of the systems formed of up to several or even
dozens of particles have been obtained~\cite{Guardiola_00,PhysRevC.56.1720,NCSM_00,cct_07}.
The progress in multi-particle
scattering problem is moderated however. The major obstacle turns to be the
rich variety of the reactions one should consider simultaneously and the
resultant complexity of the wave function asymptotic structure. Till now
only three-body system has been treated in a full extent, comprising elastic
and break-up channels~\cite{FRGL_3B,Kievsky_3B,Deltuva_3B}, whereas rigorous description
by the same methods of the four particle
scattering remains limited to the elastic and rearrangement channels~\cite{Viviani_4B,Lazauskas_4B,Deltuva_4B}.
Recently very
courageous effort has been undertaken to apply Green Function Monte Carlo~\cite{GFMC_SC_07}
and No-Core Shell model~\cite{NCSM_SC_10} calculations for the nucleon
scattering on A$\geq$4 nuclei, nevertheless these promising techniques
remains limited to the description of the binary scattering process.
Therefore a method which enable the scattering problem to be solved without explicit use of the
asymptotic form of the wave function is of great importance.

The complex
scaling method has been proposed~\cite{Nuttal_csm,CSM_71} and successfully applied for the resonance
scattering~\cite{Moiseyev}, as has been demonstrated recently this method can be extended
also for the scattering problem~\cite{CSM_Curdy_04,Elander_CSM}.
 In this study we propose novel method to solve quantum few-particle
scattering problem based on complex scaling method, which allows to use
trivial boundary conditions. We demonstrate success of this method in
both calculating elastic and three particle break-up observables.

\section{Formalism: 2-body case}

\subsubsection{Short range interaction}

The complex scaling method has been proposed a while ago to treat the
scattering problem for the exponentially
bound potentials~\cite{Nuttal_csm}. Idea is quite simple and can be summarized as follows.
First one recast the Schr\"{o}dinger equation into inhomogeneous (driven)
form by splitting systems wave function into the sum $\Psi =$ $\Psi
^{sc}+\Psi ^{in}$ containing the incident (free) $\Psi ^{in}(\mathbf{r})=\exp (i%
\mathbf{k}\cdot \mathbf{r})$ and the scattered $\Psi ^{sc}(\mathbf{r})$
waves as:
\begin{equation}
(E-\widehat{H_{0}}-V(\mathbf{r}))\Psi ^{sc}(\mathbf{r})=V(\mathbf{r})\Psi
^{in}(\mathbf{r}).  \label{Schro_2B}
\end{equation}
The scattered wave in the asymptote is represented by the outgoing
wave $\Psi ^{sc}\infty \exp (ikr)/r$. If one scale all the particle
coordinates by a constant complex factor, i.e. $\overline{r}_{i}=e^{i\theta
}r_{i\text{ }}$ with $Im(e^{i\theta })>0$, the scattered wave vanish
exponentially as $\overline{\Psi }^{sc}\infty \exp (-kr\sin \theta )$ with
increasing particle separation $r$. Moreover ff the interaction is of short range
(exponentially bound), the right hand side of eq.~(\ref{Schro_2B}) also
turns to zero for large $r$, enabling to solve former equation in a similar
way as a bound state problem: using compact basis or by solving differential
equation on a finite domain and by imposing $\overline{\Psi }^{sc}$ to vanish on
its borders.

In practice one solves 2-body problem by expanding Schr\"{o}dinger equation into partial waves:
\begin{equation}
(\frac{\hbar ^{2}}{2\mu }k^{2}-\widehat{H}_{0l}(r)-V_{l}(r))\psi
_{l}^{sc}(r)=V_{l}(r)\psi _{l}^{in}(r),
\end{equation}%
The radial part of the incoming wave is represented by the regular Bessel
functions $\psi _{l}^{in}(r)=j_{l}(kr)kr$ and\ with kinetic energy term
given by
\begin{equation}
\widehat{H}_{0l}(r)=\frac{\hbar ^{2}}{2\mu }\left[ -\frac{d^{2}}{dr^{2}}+%
\frac{l(l+1)}{r^{2}}\right] .
\end{equation}%
after the complex scaling, this equation becomes:
\begin{equation}
(\frac{\hbar ^{2}}{2\mu }k^{2}-\widehat{H}_{0l}(re^{i\theta
})-V_{l}(re^{i\theta }))\overline{\psi }_{l}^{sc}(r)=V_{l}(re^{i\theta })%
\overline{\psi }_{l}^{in}(r),
\end{equation}

The complex scaled inhomogeneous term is easily obtained by using analytical
expressions for the regular Bessel function $\overline{\psi }%
_{l}^{in}(r)=j_{l}(kre^{i\theta })kre^{i\theta }$. Extraction of the
scattering phase-shift may be done either directly by determining asymptotic
normalization coefficient of the outgoing wave:
\begin{equation}
\overline{\psi }_{l}^{sc}(r)=A_{l}(r)\exp (ikre^{i\theta }-l\pi /2)
\label{local_2B_0}
\end{equation}%
with scattering amplitude given by:
\begin{equation}
kf_{l}=e^{i\delta _{l}}\sin \delta _{l}=A_{l}(r\rightarrow \infty )
\label{local_2B_1}
\end{equation}
Other well known alternative is to use integral representation, which one gets after applying Green's
theorem:
\begin{equation}
f_{l}=-\frac{2\mu }{\hbar ^{2}}\int j_{l}(kre^{i\theta })V_{l}(re^{i\theta
})(\overline{\psi }_{l}^{sc}(r)+\overline{\psi }_{l}^{in}(r))re^{2i\theta }dr
\label{integral_2B}
\end{equation}

\subsubsection{Coulomb plus short range interaction}

If interaction contains long range term the problem turns to be quite
different. The right hand side of eq.~(\ref{Schro_2B}) after the complex
scaling diverges and the $\Psi ^{sc}(re^{i\theta })$ term is not anymore
exponentially bound. In ref.~\cite{CSM_Curdy_04,Elander_CSM} exterior
complex scaling was proposed as a
solution to circumvent the problem due to diverging term on the right hand
side of eq.~(\ref{Schro_2B}).

In this paper, when considering problem of the interaction containing short
range ($V^{s}(r)$) plus Coulomb term ($V^{C}(r)=\frac{\hbar ^{2}\eta }{\mu r}
$) we propose to stick with the standard smooth scaling procedure, however
to employ analytically continued Coulomb waves to circumvent the problem of
the divergence. In this case driven partial wave Schr\"{o}dinger equation is
written as:
\begin{equation}
(\frac{\hbar ^{2}}{2\mu }k^{2}-H_{0l}(r)-V^{s}(r)-V^{C}(r))\psi
_{l}^{sc,C}(r)=V^{s}(r)\psi _{l}^{in,C}(r),  \label{SchroC_2B}
\end{equation}%
where $\psi _{l}^{in,C}=F_{l}(\eta ,kr)$ is a regular Coulomb function,
which is solution of the former Hamiltonian containing Coulomb interaction
only. Asymptotically the scattered wave behaves as $\psi ^{sc,C}\infty \exp
(ikr-\eta \ln 2kr)$, and therefore vanish exponentially after the complex
scaling.
\begin{equation}
\overline{\psi }_{sc}^{C}(r)=A_{l}(r)\exp (ikre^{i\theta }-\eta \ln
2kre^{i\theta }-l\pi /2).  \label{local_2B_cl0}
\end{equation}

Equation~(\ref{SchroC_2B}) may be readily solved with the vanishing boundary
condition for $\overline{\psi }_{sc}^{C}(r)$, one simply must be able to continue analytically the regular
Coulomb functions standing on the right hand side.

The scattering amplitude and Coulomb corrected phase shift due to short
range interaction $\delta _{l}$ can be determined as previously from the
asymptotic normalization coefficient

\begin{equation}
e^{-i\sigma _{l}}f_{l}=e^{i(\delta _{l}+\sigma _{l})}\sin \delta
_{l}=A_{l}(r\rightarrow \infty ),  \label{local_2B_cl1}
\end{equation}%
where $\sigma _{l}$ is so-called Coulomb phase shift.

Alternatively, Green's theorem may be used to obtain integral relation
similar to eq.(\ref{integral_2B}):
\begin{equation}
f_{l}=-\frac{2\mu }{\hbar ^{2}}e^{2i\sigma _{l}}\int F_{l}(\eta
,kre^{i\theta })V^{s}(re^{i\theta })\overline{\Psi }^{C}(r)re^{2i\theta }dr
\label{integral_2B_cl}
\end{equation}

\section{Formalism: 3-body case}

\subsubsection{Short range interaction}

In sake of simplicity let consider a system of three identical spinless particles
submitted to short range pairwise interactions. Only two vector variables
are needed in the barycentric system, which may be one of the Jacobi pairs $%
\mathbf{x}_{i}=\mathbf{r}_{j}-\mathbf{r}_{k}$ and $\mathbf{y}_{i}=\frac{2}{%
\sqrt{3}}\left( \mathbf{r}_{i}-(\mathbf{r}_{k}+\mathbf{r}_{j})\right) $. The
pair potential is assumed to support any number of two-particle bound states
$\phi _{_{m}}(\mathbf{x}_{i})$ with eigenvalues $\epsilon _{m}$ and the
angular momentum of this state $l_{m}$. The corresponding continuum state
has relative momenta $q_{m}$, satisfying energy conservation relation $E=%
\frac{\hbar ^{2}}{m}q_{m}^{2}+\epsilon _{m}=K^{2};$ the second equality
defines three-particle break-up momenta $K$.

Three particle problem we formulate by using Faddeev equations~\cite{Faddeev:1960su} in
configuration space and by readily separating incoming wave of the
particle scattered on a bound pair in the state $\phi _{_{m}}(\mathbf{x}%
_{i})$. Three-identical particle problem maybe concluded in a single
equation:
\begin{equation}
(E-H_{0}-V_{i}(\mathbf{x}_{i}))\psi _{i,m}^{sc}(\mathbf{x}_{i},\mathbf{y}%
_{i})-V_{i}(\mathbf{x}_{i})\sum_{j\neq i}\psi _{j,m}^{sc}(\mathbf{x}_{j},%
\mathbf{y}_{j})=V_{i}(\mathbf{x}_{i})\sum_{j\neq i}\phi _{_{m}}(\mathbf{x}%
_{j})\exp (i\mathbf{q}_{m}\cdot \mathbf{y}_{j}), \label{FY_eq_nc}
\end{equation}%
where $\psi _{i,m}^{sc}$ is the scattered part of the Faddeev amplitude,
corresponding to the incoming particle $i$; while by $V_{i}$ we denote a pair
interaction of the particles $j$ and $k$. Decomposition of the systems wave
function into three Faddeev amplitudes permits to separate two-cluster
particle channels, whereas three-body break-up component remains shared by
the three Faddeev amplitudes. In the $y_{i}\rightarrow \infty $ asymptote
the scattered part of the Faddeev amplitude $i$ takes form
\begin{equation}
\psi _{i,m}^{sc}(\mathbf{x}_{i},\mathbf{y}_{i})\underset{y_{i}\rightarrow
\infty }{=}A_{m}(\widehat{x}_{i},\widehat{y}_{i},x_{i}/y_{i})\frac{\exp
(iK\rho )}{\rho ^{5/2}}+\sum\limits_{n}f_{nm}(\widehat{y}_{i})\phi _{_{n}}(%
\mathbf{x}_{i})\frac{\exp (iq_{n}y_{i})}{\left\vert y_{i}\right\vert }
\end{equation}%
here $\rho =\sqrt{x_{i}^{2}+y_{i}^{2}}$ is hyperradius of the system; $A_{m}(%
\widehat{x}_{i},\widehat{y}_{i},x_{i}/y_{i})$ is a three-particle break-up
amplitude; $f_{mn}(\widehat{y}_{i})$ is two-body transition amplitude from channel $m$ to channel
$n$. In this expression sum runs over all open binary channels $n$.

One may easily see that the scattered part of the Faddeev amplitude might
vanish for large hyperradius if particle coordinates are properly complex scaled
 and thus: $\overline{\mathbf{x}}_{i}=\mathbf{x}_{i}e^{i\theta },%
\overline{\mathbf{y}}_{i}=\mathbf{y}_{i}e^{i\theta }$ and $\rho =\rho
e^{i\theta }$. However in order to perform resolution of the problem on a
finite grid one should ensure that the inhomogeneous term, standing in the right hand side of
eq.~(\ref{FY_eq_nc}), also vanishes outside the resolution domain.
The inhomogeneous term is null, damped
by the potential term, if $x_{i}$ is large and thus falls outside of the interaction region.
Alternatively for $x_{i}\ll y_{i}$, the modulus of the transformed Jacobi coordinates approach $%
x_{j}\approx \frac{\sqrt{3}}{2}y_{i}$; $y_{j}\approx y_{i}/2$ and
\begin{equation}
\phi _{_{n}}(\mathbf{x}_{j}e^{i\theta })\exp (i\mathbf{q}_{n}\cdot \mathbf{y}%
_{j}e^{i\theta })\propto \phi _{n}(\mathbf{x}_{j}e^{i\theta })\frac{\exp
(-iq_{n}y_{j}e^{i\theta })}{\left\vert y_{j}\right\vert e^{i\theta}}\propto \exp(-k_{n}%
\frac{\sqrt{3}}{2}y_{i}\cos \theta)\frac{ \exp (q_{n}\frac{y_{i}}{2}\sin \theta )}{y_i},
\end{equation}%
here we explored the fact that the bound state wave function decrease exponentially
in the asymptote with momenta $k_{n}=\sqrt{m\left\vert
\epsilon _{n}\right\vert }/\hbar $. The last expression vanish for large $y_{i}$ only if condition
\begin{equation}
\tan \theta <\frac{k_{n}\sqrt{3}}{q_{n}}=\sqrt{\frac{3\left\vert \epsilon
_{n}\right\vert }{E+\left\vert \epsilon _{n}\right\vert }}\label{3b_cond_nc}
\end{equation}
is satisfied~\footnote{One must note that this condition is derived for a system of
three particles with identical masses. For the case of three particles with
arbitrary masses, one gets: $\tan \theta <\sqrt{\frac{\left\vert \epsilon
_{n}\right\vert Mm_{i}}{(E+\left\vert \epsilon _{n}\right\vert )m_{j}m_{k}}}$;
where $m_{i}$ is the mass of incoming particle, the binding energy $\left\vert \epsilon _{n}\right\vert$
 correspond the weakest bound state of
the particle pair $(jk)$ open for scattering, $M=m_{i}+m_{j}+m_{k}$ is a
total mass of the system.}. Therefore if the scattering energy is large enough one
is obliged to restrict the complex scaling angle to very small values

Extraction of the scattering observables maybe realized as previously in two
different ways. Straightforward way is to extract transition amplitudes from
the $y_{i}\rightarrow \infty $ asymptote of the solution $\overline{\psi }%
_{i}^{sc}(\mathbf{x}_{i},\mathbf{y}_{i}),$ exploiting the fact that
different scattering channels are mutually orthogonal:
\begin{equation}
f_{nm}(\widehat{y}_{i})=C_{n}^{-1}\left\vert y_{i}\right\vert \exp
(-iq_{n}y_{i}e^{i\theta })\int \phi _{_{n}}^{\ast }(\mathbf{x}%
_{i}e^{-i\theta })\overline{\psi }_{i,m}^{sc}(\mathbf{x}_{i},\mathbf{y}%
_{i})e^{3i\theta }d^{3}\mathbf{x}_{i,} \label{3b_amp_nc}
\end{equation}%
where $C_{n}$ is normalization coefficient of two-body wave function $\int
\phi _{_{n}}^{\ast }(\mathbf{x}_{i}e^{-i\theta })\phi _{_{n}}(\mathbf{x}%
_{i}e^{i\theta })e^{3i\theta }d^{3}\mathbf{x}_{i}=C_{n}$. Break-up amplitude might be
extracted from the $\overline{\psi }_{i}^{sc}(\mathbf{x}_{i},\mathbf{y}_{i})$
once all the two-body transition amplitudes are calculated relaying on eq.~(\ref{3b_amp_nc}).

Alternatively one can employ Green's theorem. In this case integral relations
might be obtained both for break-up and two-body transition amplitudes. For
the transition amplitude one has:
\begin{equation}
f_{nm}(\widehat{y}_{i})=-C_{n}^{-1}\frac{m}{\hbar ^{2}}\int \int \phi
_{_{n}}^{\ast }(\mathbf{x}_{i}e^{-i\theta })\frac{\exp
(-iq_{n}y_{i}e^{i\theta })}{\left\vert y_{i}\right\vert }(V_{j}(\mathbf{x}%
_{j}e^{i\theta })+V_{k}(\mathbf{x}_{k}e^{i\theta }))\overline{\Psi }_{m}(%
\mathbf{x}_{i},\mathbf{y}_{i})e^{6i\theta }d^{3}\mathbf{x}_{i}d^{3}\mathbf{y}%
_{i} \label{3b_amp_int}
\end{equation}
 These integrals are convergent on the finite domain, if the
following condition is satisfied:
\begin{equation}
\tan \theta <\frac{\sqrt{3}k_{m}}{q_{m}+2q_{n}}=\frac{\sqrt{3\left\vert
\epsilon _{m}\right\vert }}{\sqrt{\left\vert \epsilon _{m}\right\vert +E}+2%
\sqrt{\left\vert \epsilon _{n}\right\vert +E}},
 \label{3b_cond_strong}
\end{equation}%
which is stronger than condition eq.~(\ref{3b_cond_nc})~\footnote{For the case of three particles with
arbitrary masses, one gets: $\tan \theta <\sqrt{m_{i}M\left\vert \epsilon
_{m}\right\vert }/(\sqrt{m_{j}m_{k}(\left\vert \epsilon _{m}\right\vert +E)}+%
\sqrt{(M-m_{j})(M-m_{k})(\left\vert \epsilon _{n}\right\vert +E)})$}.

For the break-up amplitude several different relations can be obtained~\cite{Glockle_book}, it
seems that the one employing 2-body outgoing states $\overline{\phi }%
^{(+)}(p,\mathbf{x}_{i})$, generated by the correspondingly scaled strong
potential at relative momenta $p$, seems the most reliable numerically:
\begin{equation}
A_{i,m}(\widehat{x}_{i},\widehat{y}_{i},x_{i}/y_{i})=\frac{m}{\hbar ^{2}}%
\int \int \overline{\phi }^{(+)}(Kx_{i}/y_{i},\mathbf{x}_{i})\frac{\exp
(-iq_{n}y_{i}e^{i\theta })}{\left\vert y_{i}\right\vert }V_{i}(\mathbf{x}%
_{i}e^{i\theta })(\overline{\psi }_{j,m}+\overline{\psi }_{k,m})e^{6i\theta
}d^{3}\mathbf{x}_{i}d^{3}\mathbf{y}_{i}
\label{3b_bramp_int}
\end{equation}

In practice, calculations are performed by expanding former equations into
partial waves. This pure technical issue is not subject of this paper and we
refer interested reader to~\cite{Glockle_book} for the details on the partial wave Faddeev
equations. One should note that partial wave expansion has no effect on the
validity of the presented method.

\subsubsection{\protect\bigskip Short range interaction plus Coulomb}

The former discussion can be readily extended for the case when particles
interact via short range plus Coulomb forces. In this case we prefer to use
Faddeev-Merkuriev equations~\cite{Stas_80}, which elaborate Faddeev
formalism by accommodating Coulomb force. Indeed, Faddeev equations,
suppose free asymptotic behavior of the particles, in case of long range
interaction become ill behaved due to noncompactness of their kernel. These
equations can still provide satisfactory solution for bound state problem,
but are impractical for the scattering case~\footnote{In principle,
Faddeev equations maybe solved also for long-range scattering
problem after the complex scaling, however only vanishing
of the total wave function and not of the separate Faddeev
amplitudes (ones work with) is assured in this case.}.
Faddeev equations does not shed light on the asymtotic behavior of the separate
amplitudes when long range interaction is present. Idea of
Merkuriev~\cite{Stas_80} is to split the Coulomb potential $V^{C}$ into two parts
(short and long range), $V^{C}=V^{s.C}+V^{l.C}$, by means of some arbitrary
cut-off function $\chi $:
\begin{equation}
V_{i}^{s.C}(x_{i},y_{i})=V_{i}^{C}(x_{i})\chi _{i}(x_{i},y_{i})\qquad
V_{i}^{l.C}(x_{i},y_{i})=V_{i}^{C}(x_{i})[1-\chi _{i}(x_{i},y_{i})]
\end{equation}%
and reshuffle long range terms. One is then left with a system of equivalent
equations:
\begin{equation}
(E-H_{0}-W_{i}-V_{i}^{s})\psi _{i}^{sc}-V_{i}^{s}\sum_{i\neq j}(\psi
_{j}^{sc}+\psi _{j,m}^{in})=(W_{i}-V_{i}^{l}-V_{i}^{C.res})\psi
_{i,m}^{in}\qquad W_{i}=V_{i}^{l}+V_{j}^{l}+V_{k}^{l}
\end{equation}%
by $V_{i}^{s}$ we consider sum of the short range interaction $V_{i}(x_{i})$
of the pair $(jk)$ plus short range part of the Coulomb force $%
V_{i}^{s}=V_{i}^{s.C}+V_{i}$. The incoming state is defined by:
\begin{equation}
\psi _{i,m}^{in}(\mathbf{x}_{i},\mathbf{y}_{i})=\phi _{_{m}}(\mathbf{x}%
_{i})\varphi ^{C}(q_{m},\mathbf{y}_{i})
\end{equation}%
where $\varphi ^{C}(q_{m},\mathbf{y}_{i})$ is a plane wave of the incident
particle $i$ moderated by its residual Coulomb interaction with a cluster of
particles $(jk)$: $V_{i}^{C.res}(y_{i})=$ $\frac{4\hbar ^{2}\eta _{i,jk}}{%
\sqrt{3}my_{i}}$. The $\phi _{_{m}}(\mathbf{x}_{i})$ is the eigenfunction of
the $m-$th bound state of the particle pair $(jk)$.

Former equations are solved as well as the scattering observables extracted
in a similar way as for the Coulomb free case. For example transition amplitude, via Green's theorem,
is expressed as:
\begin{eqnarray}
f^C_{nm}(\widehat{y}_{i})=-C_{n}^{-1}\frac{m}{\hbar ^{2}}\int \int \phi
_{_{n}}^{\ast }(\mathbf{x}_{i}e^{-i\theta })\varphi^{C*}(q_m,\mathbf{y}_ie^{-i\theta })\times\nonumber \\
(V^s_{j}(\mathbf{x}%
_{j}e^{i\theta })+V^s_{k}(\mathbf{x}_{k}e^{i\theta })+W_i(\mathbf{x}_ie^{i\theta },\mathbf{y}_ie^{i\theta })-
V^{C.res}( y_{i}e^{i\theta })-V_i(x_ie^{i\theta }))
\overline{\Psi }_{m}(\mathbf{x}_{i},\mathbf{y}_{i})e^{6i\theta }d^{3}\mathbf{x}_{i}d^{3}\mathbf{y},%
_{i} \label{3b_amp_int}
\end{eqnarray}
with $V_i(x_i)$ representing full interaction between particle-pair (jk). 

There is however formal
difficulty associated with the extraction of the break-up amplitude for the
case when all three particles are charged, since the asymptotic form of the
break-up wave function is not known. One may still rely on the approximate
relation employing Peterkop integral~\cite{Peterkop} as is claimed in~\cite{PhysRevLett.93.233201}.

\section{Results}

To test our approach we consider model of nucleons with mass $\frac{\hbar
^{2}}{m}=41.47$ $MeV\cdot fm^{2}$, where strong part of nucleon-nucleon (NN)
interaction is described by S-wave MT I-III potential, defined as:
\begin{equation}
V_{S}(r)=(A_{S}\exp (-1.55r)+1438.72\exp (-3.11r))/r
\end{equation}%
where $V_{S}(r)$ is in MeV and $r$ is in fm units. The attractive Yukawa
term is defined with $A_{s=0}=-513.968$ $MeV\cdot fm$ and $A_{s=1}=-626.885$
$MeV\cdot fm$ for the two-nucleon interaction in spin singlet and triplet
states respectively.

 MT I-III potential has been chosen for two reasons. On one hand it is widely
employed potential for which accurate benchmark calculations exist.
On the other hand this potential, being a combination of the attractive and
repulsive Yukawa terms, reflects well the structure of the realistic
nucleon-nucleon interaction: it is strongly repulsive at the origin, however
has narrow attractive well situated at $r\approx 1$ fm. Note that many
numerical techniques fail for the potentials, like MT I-III, which contain a
repulsive core.

\begin{figure}[!]
\begin{center}
\mbox{\epsfxsize=8.5cm\epsffile{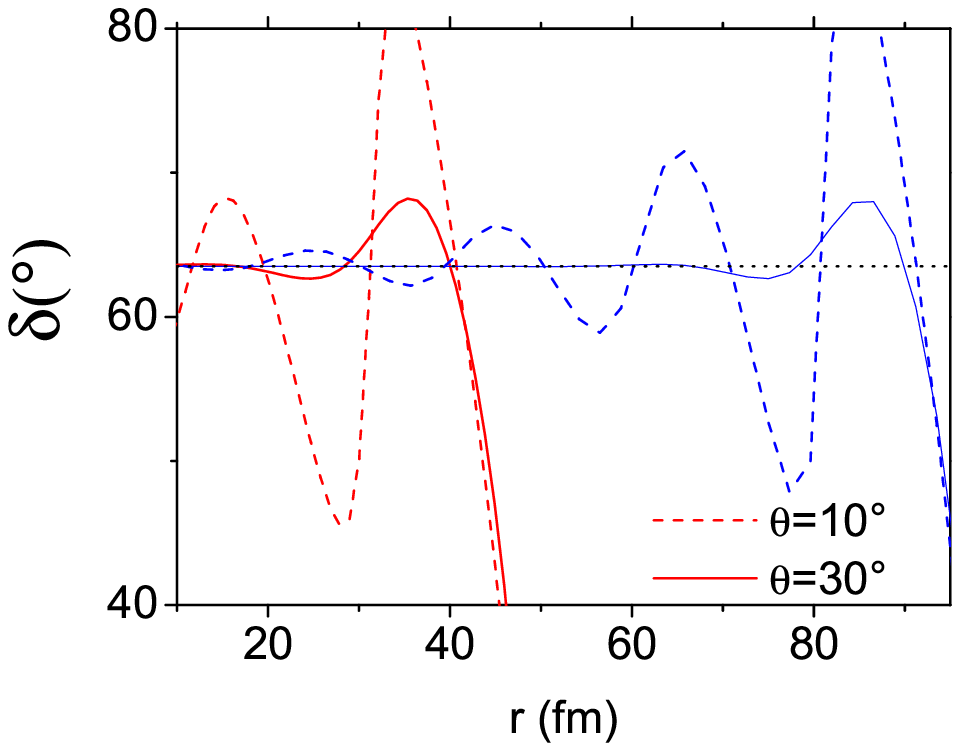}} \mbox{\epsfxsize=8.5cm\epsffile{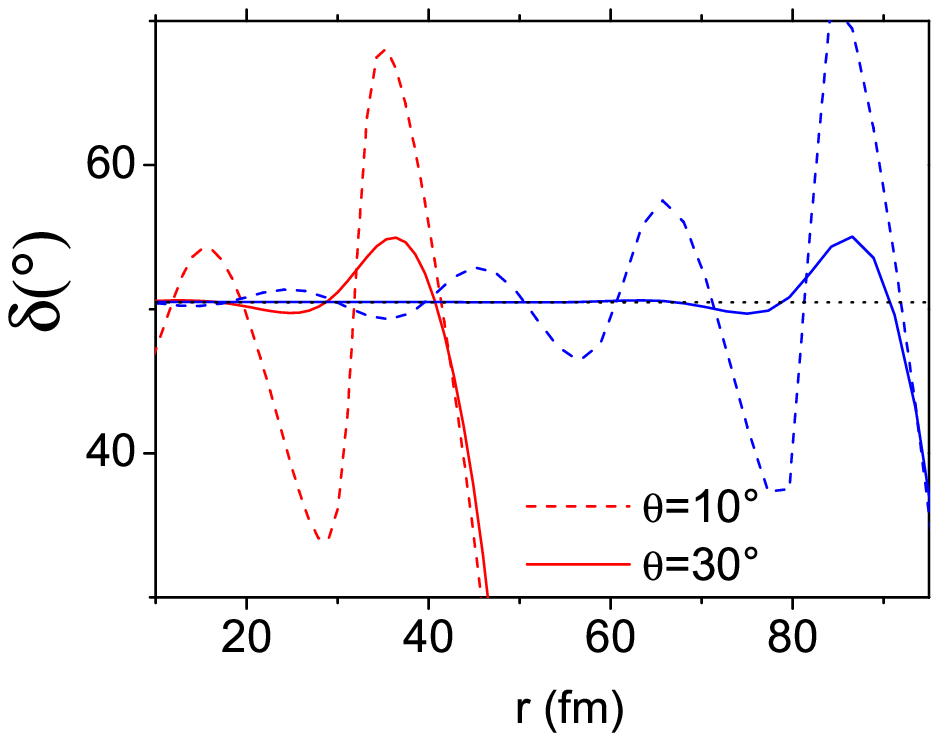}}
\end{center}
\caption{(Color online) $^1$S$_0$ NN phaseshift at $E_{cm}$=1 MeV extracted using relations
eq.~\protect\ref{local_2B_0}-\protect\ref{local_2B_1} and ~\protect\ref%
{local_2B_cl0}-\protect\ref{local_2B_cl1} respectively. Calculations
performed with cut-off imposed at $r_{max}$=50 and 100 fm using complex
rotation by the angle $\protect\theta$ =10$^\circ$ (dashed lines) and $%
\protect\theta =$30$^\circ$ (solid line). Pure strong interaction result is
presented in the left figure, calculations including repulsive Coulomb
interaction for pp-pair are presented in the right figure.}
\label{fig:r_dep}
\end{figure}

In figure~\ref{fig:r_dep} we present our calculations for the NN $^{1}$S$%
_{0} $ phaseshift at $E_{cm}=$1 MeV. Two calculation sequences have been
performed by enforcing the $\overline{\psi }_{l}^{sc}(r)$ to vanish at the
border of the numerical grid set at 50 fm and at 100 fm respectively, whereas complex
scaling angle $(\theta )$ has been chosen to be 10$^\circ$ and 30$^\circ$.
The phaseshift is extracted from the $\overline{\psi }_{l}^{sc}(r)$ value at
fixed distance $r$, according to eq.~\ref{local_2B_0}-\ref{local_2B_1}
(Coulomb free case) and eq.~\ref{local_2B_cl0}-\ref{local_2B_cl1} (short
range plus Coulomb interaction). Extracted phaseshift oscillates with
$r$ -- this oscillatory behavior is due to the "premature" enforcement of
$\overline{\psi }_{l}^{sc}(r)$ to vanish at the border of the grid $r_{\max }$.
The phaseshifts extracted close to $r_{\max }$ are strongly affected by the cut-off
and thus not reliable. The amplitude of the close-border oscillations
is sizeably reduced by either increasing $r_{\max }$ or $\theta $, i.e. by reducing
the sharpness of the numerical cut-off. Extracted phaseshift from the calculation with $%
r_{\max }=100$ fm and $\theta =$30$^\circ$ is stable in a rather large window,
which starts at $r\sim 5$ fm (right outside the interaction region) and
extends up to $r\sim 70$ fm (beyond this value effect due to cut-off sets
in). In the stability region extracted phaseshift value agrees well with the exact result
(dotted line).

\begin{figure}[!]
\begin{center}
\mbox{\epsfxsize=8.5cm\epsffile{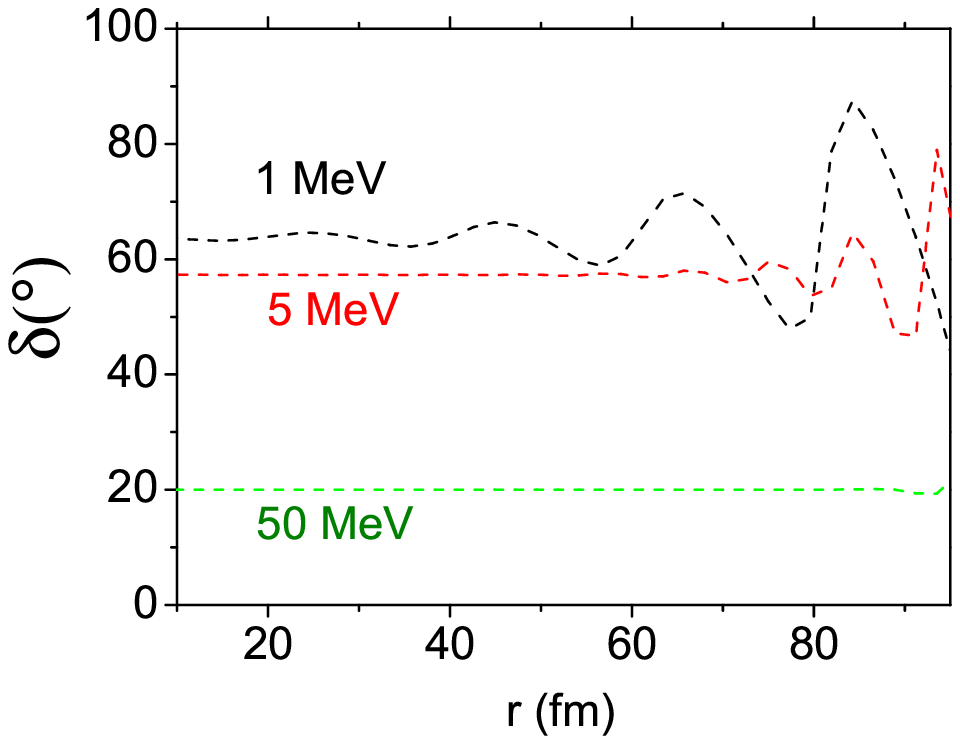}} \mbox{\epsfxsize=8.5cm%
\epsffile{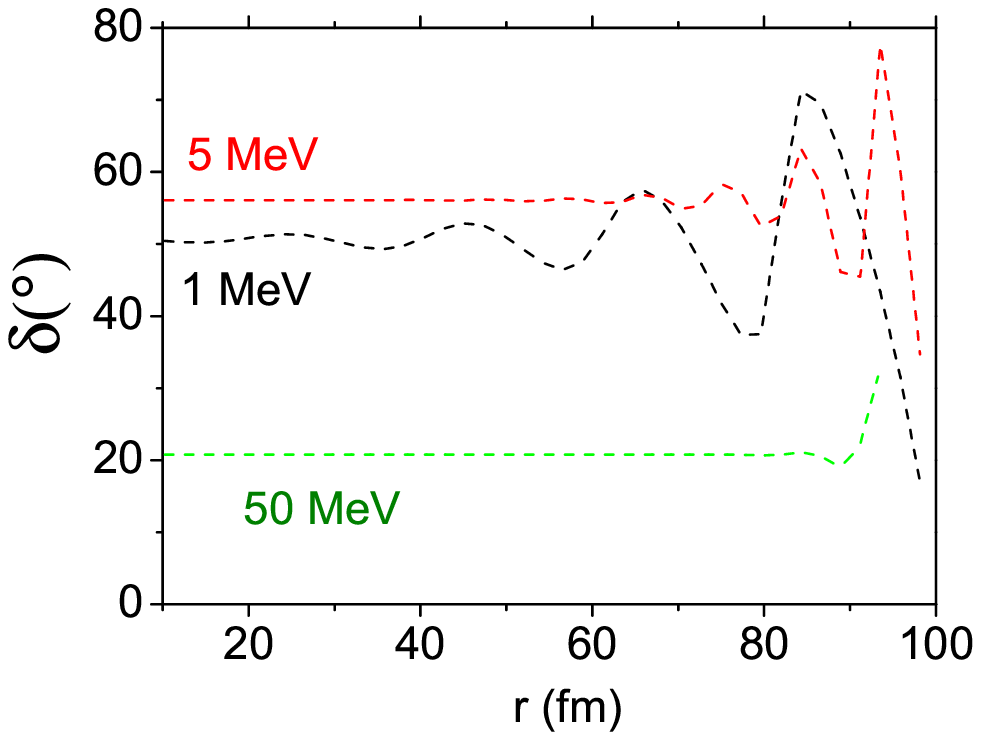}}
\end{center}
\caption{(Color online) $^1$S$_0$ NN phaseshift calculation at $E_{cm}$=1, 5 and 50 MeV
extracted using relations eq.~\protect\ref{local_2B_0}-\protect\ref%
{local_2B_1} and ~\protect\ref{local_2B_cl0}-\protect\ref{local_2B_cl1}
respectively. Calculations performed with cut-off imposed at $r_{max}$= 100
fm using complex rotation by the angle $\protect\theta$ =10$^\circ$. Pure
strong interaction result is presented in the left figure, calculations
including repulsive Coulomb interaction for pp-pair are presented in the
right figure.}
\label{fig:r_dep_en}
\end{figure}

In figure~\ref{fig:r_dep_en} we compare NN $^{1}$S$_{0}$ pahaseshift
calculation for $E_{cm}$=$1$, $5$ and $50$ MeV by setting $r_{\max }=100$ fm
and $\theta$ =10$^\circ$. One may see that by increasing energy effect of
the cut-off reduces, sizeably improving stability of the extracted
phaseshift. Inclusion of the repulsive Coulomb term does not have any effect
on the quality of the method.

\begin{figure}[tbp]
\begin{center}
\mbox{\epsfxsize=8.0cm\epsffile{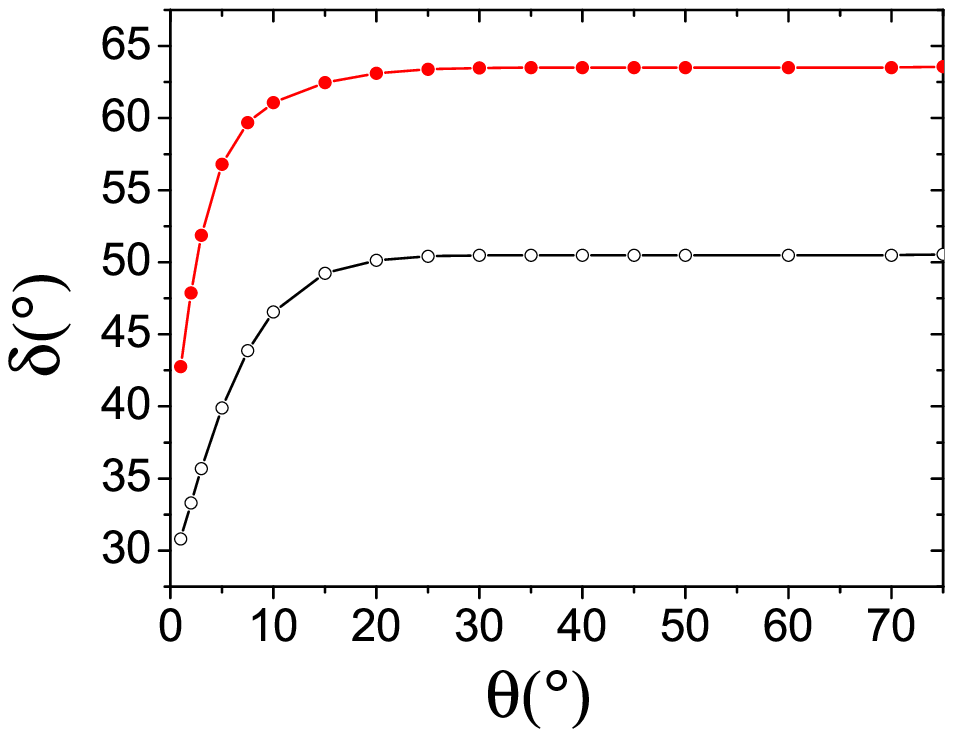}}\hspace{1.cm}
\end{center}
\caption{(Color online) Dependence of the calculated NN  $^1$S$_0$ phaseshift using integral
expression as a function of the complex rotation angle. Grid was limited to r%
$_{max}$=100 fm. The upper curve correspond Coulomb-free case, the bottom
one includes Coulomb.}
\label{fig:integral_2B}
\end{figure}

One may improve considerably accuracy of the phaseshift calculation by
employing integral relation eq.~(\ref{integral_2B}), see tables~\ref%
{tab_2B_1M},\ref{tab_2B_50M} and figure~\ref{fig:integral_2B}.
Phaseshift converges to constant value by either  increasing cut-off radius r$%
_{\max }$ or complex rotation angle. Accuracy of five digits
is easily reached. One should notice however that the
use of very large values of $\theta$ should be avoided, due to the fact
that the function $\overline{\psi }_{l}^{sc}(r)$ as well as complex scaled potential $V(re^{i\theta})$
might become very steep and rapidly oscillating.
At higher energy the function  $\overline{\psi }_{l}^{sc}(r)$ vanishes faster
and thus one achieves convergence by employing smaller values of r$_{\max }$ and/or $\theta$.

\begin{table}[h!]
\caption{Calculation of the scattering phaseshift using integral expressions
at $E_{cm}=$1 MeV }
\label{tab_2B_1M}%
\begin{ruledtabular}
\begin{tabular}{c|cccc|cccc|}
& \multicolumn{4}{c|}{MT I-III} & \multicolumn{4}{c|}{MT I-III+Coulomb}\\\hline
$r_{max}$ (fm)& 5$^\circ$ & 10$^\circ$ & 30$^\circ$ & 50$^\circ$ & 5$^\circ$ & 10$^\circ$ & 30$^\circ$ & 50$^\circ$ \\\hline
10 & 44.420 & 49.486 & 55.790 & 56.676 & 33.999 & 36.390 & 41.528 & 43.805
\\
25 & 34.704 & 44.211 & 62.654 & 63.743 & 24.772 & 34.910 & 50.693 & 50.698
\\
50 & 56.812 & 61.083 & 63.482 & 63.512 & 39.895 & 46.546 & 50.487 & 50.491
\\
100 & 66.502 & 63.822 & 63.512 & 63.512 & 55.463 & 50.811 & 50.491 & 50.491
\\
150 & 62.497 & 63.485 & 63.512 & 63.512 & 49.317 & 50.474 & 50.491 & 50.491
\\
\hline
exact & \multicolumn{4}{c|}{63.512} & \multicolumn{4}{c|}{50.491}\\
\end{tabular}
\end{ruledtabular}
\end{table}

\begin{table}[h!]
\caption{Calculation of the scattering phaseshift using integral expressions
at $E_{cm}=$50 MeV }
\label{tab_2B_50M}%
\begin{ruledtabular}
\begin{tabular}{c|cccc|cccc|}
$r_{max}$ (fm)& \multicolumn{4}{c|}{MT I-III} & \multicolumn{4}{c|}{MT I-III+Coulomb}\\\hline

& 3$^\circ$ & 5$^\circ$ & 10$^\circ$ & 30$^\circ$ & 3$^\circ$ & 5$^\circ$ & 10$^\circ$ & 30$^\circ$ \\\hline
10 & 19.400 & 19.719 & 19.923 & 19.605 & 19.795 & 20.245 & 20.610 & 20.313
\\
25 & 20.788 & 20.135 & 20.027 & 20.032 & 21.530 & 20.864 & 20.755 & 20.760
\\
50 & 20.014 & 20.026 & 20.027 & 20.027 & 20.734 & 20.754 & 20.755 & 20.755
\\
100 & 20.027 & 20.027 & 20.027 & 20.027 & 20.755 & 20.755 & 20.755 & 20.755
\\\hline
exact & \multicolumn{4}{c|}{20.027} & \multicolumn{4}{c|}{20.755}\\
\end{tabular}
\end{ruledtabular}
\end{table}

\bigskip

Our analysis has been extended to the three-body case.
We consider nucleon-deuteron (N-d) $L=0$ scattering
in spin-doublet ($S=1/2$) and quartet ($S=3/2$) states.
Calculations has been performed both below and above three-particle
break-up threshold. Below the break-up threshold results are stable
and independent of the scaling angle, similar to the two-body case. Phaseshifts might
be accurately extracted both using differential and integral expressions.

Application of the differential relations for extracting scattering phaseshift and inelasticity
above the break-up threshold is not so-obvious. It is difficult to find the stability domain.
Therefore we employed
integral expressions eqs. (\ref{3b_amp_int}) and (\ref{3b_bramp_int}), obtained using Greens theorem, which once again proved their worth.
We summarize obtained
results in Table~\ref{tab_3Bnd_brca}
and ~\ref{tab_3Bpd_brca}, respectively for n-d and p-d scattering above the break-up threshold.
Very accurate results are obtained for both phaseshift and inelasticity parameter
once complex scaling angle is chosen in the interval $[4^\circ,12.5^\circ]$ for incident neutron with energy E$_{lab}$=14.1 MeV and $[3^\circ,7.5^\circ]$ at E$_{lab}$=42 MeV. Stability of the final result within at least three significant digits
is assured, providing excellent agreement with the benchmark calculations of~\cite{Friar_nd_bench,Arnas_bench}.
Calculated integral gradually cease to converge
on the finite domain for the calculations when higher complex scaling angles are chosen.
This happens due to the fact that condition eq.(\ref{3b_cond_strong}), which set limit
$\theta_{max}$=14.2$^\circ$ and $\theta_{max}$=8.9$^\circ$ for the calculations at E$_{lab}$=14.1 MeV
and  E$_{lab}$=42 MeV respectively, is violated.

In the table~\ref{tab_3B_brca} we tabulate $^3S_1$ n-d break-up amplitude as a function
of the break-up angle $\vartheta$, which defines pair and spectator wave numbers via k=K cos($\vartheta$) and q= 2K sin($\vartheta$)/$\sqrt{3}$ respectively.
Nice agreement is obtained with the benchmark calculation of~\cite{Friar_nd_bench}.
Small discrepancy appears only for the $\vartheta$ values close to 90$^\circ$, which
defines configuration where one pair of particles after the break-up remains at rest.
This is due to the slow convergence of the integral eq.(\ref{3b_bramp_int}) for $\vartheta\rightarrow90^\circ$ in y-direction, special
procedure must be undertaken in this particular case to evaluate the part of the slowly convergent
integral outside the resolution domain defined by y$_{max}$.

\begin{table}[tbp]
\caption{Neutron-deuteron scattering phaseshift and inelasticity parameter as a function of
the complex rotation angle $\protect\theta $ compared with benchmark
results of~\cite{Friar_nd_bench,Arnas_bench}. Our calculations has been performed by setting y$_{max}$=100 fm.}
\label{tab_3Bnd_brca}%
\begin{ruledtabular}
\begin{tabular}{ccccccccc}
& 3$^\circ$ & 4$^\circ$ & 5$^\circ$ & 6$^\circ$ & 7.5$^\circ$ & 10$^\circ$ & 12.5$^\circ$ & Ref.~\cite{Friar_nd_bench,Arnas_bench} \\\hline
 \multicolumn{9}{c}{nd doublet at E$_{lab}$=14.1 MeV } \\\hline
Re($\delta )$ & 105.00 & 105.43 & 105.50 & 105.50 & 105.50 & 105.49 & 105.48
& 105.49 \\
$\eta $ & 0.4559 & 0.4638 & 0.4653 & 0.4654 & 0.4653 & 0.4650 & 0.4649 &
0.4649 \\\hline
 \multicolumn{9}{c}{nd doublet at E$_{lab}$=42 MeV } \\\hline
Re($\delta )$ & 41.71 & 41.63 & 41.55 & 41.51 & 41.45 & 41.04 &  & 41.35 \\
$\eta $ & 0.5017 & 0.5015 & 0.5014 & 0.5014 & 0.5015 & 0.5048 &  & 0.5022 \\\hline
 \multicolumn{9}{c}{nd quartet at E$_{lab}$=14.1 MeV } \\\hline
Re($\delta )$ & 68.47 & 68.90 & 68.97 & 68.97 & 68.97 & 68.97 & 68.97 & 68.95
\\
$\eta $ & 0.9661 & 0.9762 & 0.9782 & 0.9784 & 0.9783 & 0.9782 & 0.9780 &
0.9782 \\\hline
 \multicolumn{9}{c}{nd quartet at E$_{lab}$=42 MeV } \\\hline
Re($\delta )$ & 37.83 & 37.80 & 37.77 & 37.77 & 37.74 & 38.06 & - & 37.71 \\
$\eta $ & 0.9038 & 0.9034 & 0.9032 & 0.9030 & 0.9029 & 0.8980 & - & 0.9033
\\
\end{tabular}
\end{ruledtabular}
\end{table}

\begin{table}[tbp]
\caption{Proton-deuteron scattering phaseshift and inelasticity parameter as a function of
the complex rotation angle $\protect\theta $ compared with benchmark
values of~\cite{Arnas_bench}. Our calculations has been performed by setting y$_{max}$=150 fm. }
\label{tab_3Bpd_brca}%
\begin{ruledtabular}
\begin{tabular}{ccccccccc}
& 3$^\circ$ & 4$^\circ$ & 5$^\circ$ & 6$^\circ$ & 7.5$^\circ$ & 10$^\circ$ & 12.5$^\circ$ & Ref.~\cite{Arnas_bench} \\\hline
 \multicolumn{9}{c}{pd doublet at E$_{lab}$=14.1 MeV } \\\hline
Re($\delta )$ & 108.46 & 108.43 & 108.43 & 108.43 & 108.43 & 108.43 & 108.42
& 108.41[3] \\
$\eta $ & 0.5003 & 0.4993 & 0.4990 & 0.4988 & 0.4986 &  0.4984 & 0.4981  &
0.4983[1] \\\hline
 \multicolumn{9}{c}{pd doublet at E$_{lab}$=42 MeV } \\\hline
Re($\delta )$ & 43.98 & 43.92 & 43.87 & 43.82 & 43.78 & 44.83 & - & 43.68[2]
\\
$\eta $ & 0.5066 & 0.5060 & 0.5056 & 0.5054 & 0.5052 & 0.5488 & - & 0.5056\\\hline
 \multicolumn{9}{c}{pd quartet at E$_{lab}$=14.1 MeV } \\\hline
Re($\delta )$ & 72.70 & 72.65 & 72.65 & 72.64 & 72.64 & 72.63 & 72.62 & 72.60
\\
$\eta $ & 0.9842 & 0.9827 & 0.9826 & 0.9826 & 0.9826 & 0.9828 & 0.9829 &
0.9795[1] \\\hline
 \multicolumn{9}{c}{pd quartet at E$_{lab}$=42 MeV } \\\hline
Re($\delta )$ & 40.13 & 40.11 & 40.08 & 40.07 & 40.05 & 40.35 & - & 39.96[1]
\\
$\eta $ & 0.9052 & 0.9044 & 0.9039 & 0.9036 & 0.9034 & 0.9026 & - & 0.9046
\\\hline

\end{tabular}
\end{ruledtabular}
\end{table}
\bigskip
\begin{table}[tbp]
\caption{Neutron-deuteron $^3S_1$ break-up amplitude calculated at $E_{lab}$=42 MeV as a function
of the break-up angle $\vartheta$.}
\label{tab_3B_brca}%
\begin{ruledtabular}
\begin{tabular}{ccccccccccc}
& 0$^\circ$  & 10$^\circ$  & 20$^\circ$  & 30$^\circ$   & 40$^\circ$   & 50$^\circ$  & 60$^\circ$  & 70$^\circ$  & 80$^\circ$  & 90$^\circ$    \\\hline
This work Re($^{3}$S$_{1}$) & 1.49[-2] & 8.84[-4] & -3.40[-2] & 3.33[-2] & 7.70[-2] &
2.52[-1] & 4.47[-1] & 6.47[-1] & 6.30[-1] & -1.62[-1]   \\
This work Im($^{3}$S$_{1}$) & 1.69[0] & 1.74[0] & 1.87[0] & 1.92[0] & 1.80[0] & 1.68[0]
& 1.70[0] & 1.96[0] & 2.23[0] & 3.17[0]\\\hline
Ref.~\cite{Friar_nd_bench} Re($^{3}$S$_{1}$) & 1.48[-2] & 9.22[-4] & -3.21[-2] & 3.09[-2] & 7.70[-2] &
2.52[-1] & 4.51[-1] & 6.53[-1] & 6.93[-1] & -1.05[-1]   \\
Ref.~\cite{Friar_nd_bench} Im($^{3}$S$_{1}$) & 1.69[0] & 1.74[0] & 1.87[0] & 1.92[0] & 1.80[0] & 1.67[0]
& 1.70[0] & 1.95[0] & 2.52[0] & 3.06[0]
\end{tabular}
\end{ruledtabular}
\end{table}

\section{\protect\bigskip Conclusion}

In this work we have presented a method based on the complex scaling, which enables to solve few-nucleon scattering problem without explicit treatment of the boundary conditions using square-integrable functions. Validity of the method is demonstrated for two and three particle scattering, including the three-particle break-up case with repulsive Coulomb interaction. Three-digit accuracy maybe easily obtained using this method.

As is well known for two body case complex scaling angle is, in principle, only limited to 90$^\circ$. On the contrary in order to solve three-body break-up problem the scaling angle should be restricted stronger from above, see eq.(\ref{3b_cond_strong}).
I.e. for the scattering at high energy the scaling angle should be limited to very small values. Nevertheless this
limitation does not spoil the method at high energies, since rapid vanishing of the outgoing wave after the scaling is
ensured by the large wave number values.

\begin{acknowledgments}
The authors are indebted to  A.~Deltuva for the help in benchmarking results of this manuscript
and for the fruitful discussions.
This work was granted access to the HPC resources of IDRIS under the allocation 2009-i2009056006
made by GENCI (Grand Equipement National de Calcul Intensif). We thank the staff members of the IDRIS for their constant help.
\end{acknowledgments}

\bibliographystyle{apsrev4-1}
\bibliography{cmpscale}

\end{document}